\documentclass[%
 aip,
 amsmath,amssymb,
 reprint,%
]{revtex4-2}

\usepackage{graphicx}
\usepackage{dcolumn}
\usepackage{bm}

\usepackage[utf8]{inputenc}
\usepackage[T1]{fontenc}
\usepackage{mathptmx}
\usepackage{etoolbox}
\usepackage{hyperref}

\makeatletter
\def\@email#1#2{%
 \endgroup
 \patchcmd{\titleblock@produce}
  {\frontmatter@RRAPformat}
  {\frontmatter@RRAPformat{\produce@RRAP{*#1\href{mailto:#2}{#2}}}\frontmatter@RRAPformat}
  {}{}
}%
\makeatother
\begin{document}

\preprint{AIP/123-QED}

\title[]{
Correlating Superconducting Qubit Performance Losses to Sidewall Near-Field Scattering via Terahertz Nanophotonics}
\author{Richard H. J. Kim}
\affiliation{Ames National Laboratory, US Department of Energy, Ames, IA 50011, USA}%
\author{Samuel J. Haeuser}%
\affiliation{Department of Physics and Astronomy, Iowa State University, Ames, IA 50011, USA}%
\author{Joong-Mok Park}
\affiliation{Ames National Laboratory, US Department of Energy, Ames, IA 50011, USA}%
\author{Randall K. Chan}%
\affiliation{Department of Physics and Astronomy, Iowa State University, Ames, IA 50011, USA}%
\author{Jin-Su Oh}
\affiliation{Ames National Laboratory, US Department of Energy, Ames, IA 50011, USA}%
\author{Thomas Koschny}
\affiliation{Ames National Laboratory, US Department of Energy, Ames, IA 50011, USA}%
\author{Lin Zhou}
\affiliation{Ames National Laboratory, US Department of Energy, Ames, IA 50011, USA}%
\affiliation{Department of Physics and Astronomy, Iowa State University, Ames, IA 50011, USA}%
\author{Matthew J. Kramer}
\affiliation{Ames National Laboratory, US Department of Energy, Ames, IA 50011, USA}%
\author{Akshay A. Murthy}
\affiliation{Superconducting Quantum Materials and Systems Division, Fermi National Accelerator Laboratory (FNAL), Batavia, IL 60510, USA}%
\author{Mustafa Bal}
\affiliation{Superconducting Quantum Materials and Systems Division, Fermi National Accelerator Laboratory (FNAL), Batavia, IL 60510, USA}%
\author{Francesco Crisa}
\affiliation{Superconducting Quantum Materials and Systems Division, Fermi National Accelerator Laboratory (FNAL), Batavia, IL 60510, USA}%
\author{Sabrina Garattoni}
\affiliation{Superconducting Quantum Materials and Systems Division, Fermi National Accelerator Laboratory (FNAL), Batavia, IL 60510, USA}%
\author{Shaojiang Zhu}
\affiliation{Superconducting Quantum Materials and Systems Division, Fermi National Accelerator Laboratory (FNAL), Batavia, IL 60510, USA}%
\author{Andrei Lunin}
\affiliation{Superconducting Quantum Materials and Systems Division, Fermi National Accelerator Laboratory (FNAL), Batavia, IL 60510, USA}%
\author{David Olaya}
\affiliation{National Institute of Standards and Technology, Boulder, CO 80305, USA}%
\author{Peter Hopkins}
\affiliation{National Institute of Standards and Technology, Boulder, CO 80305, USA}%
\author{Alex Romanenko}
\affiliation{Superconducting Quantum Materials and Systems Division, Fermi National Accelerator Laboratory (FNAL), Batavia, IL 60510, USA}%
\author{Anna Grassellino}
\affiliation{Superconducting Quantum Materials and Systems Division, Fermi National Accelerator Laboratory (FNAL), Batavia, IL 60510, USA}%
\author{Jigang Wang}
\affiliation{Ames National Laboratory, US Department of Energy, Ames, IA 50011, USA}%
\affiliation{Department of Physics and Astronomy, Iowa State University, Ames, IA 50011, USA}%
\date{\today}

\begin{abstract}
Elucidating dielectric losses, structural heterogeneity, and interface imperfections is critical for improving coherence in superconducting qubits. However, most diagnostics rely on destructive electron microscopy or low-throughput millikelvin quantum measurements. Here, we demonstrate noninvasive terahertz (THz) nano-imaging/-spectroscopy of encapsulated niobium transmon qubits, revealing sidewall near-field scattering that correlates with qubit coherence. We further employ a THz hyperspectral line scan to probe dielectric responses and field participation at Al junction interfaces. These findings highlight the promise of THz near-field methods as a high-throughput proxy characterization tool for guiding material selection and optimizing processing protocols to improve qubit and quantum circuit performance.
\end{abstract}


\maketitle

While superconducting qubits have emerged as a leading technology platform for quantum computing, single qubit coherence still continues to be a major limiting factor in building scalable quantum systems. With niobium (Nb) being employed as the primary material in  superconducting qubits, the formation of surface oxides on Nb has been identified as the major source of microwave loss by hosting two-level systems (TLS) \cite{ab:2024,baf:2024,oh1:2024,oh2:2024,n7}. Recent studies have demonstrated that encapsulating the surface of Nb with other materials can prevent the formation of its lossy surface oxide, thereby achieving a significant improvement in the $T_{1}$ relaxation times \cite{bal:2024}. From a materials-level standpoint, a comprehensive investigation that observes potential structural defects produced along fabrication processes in the capping approach becomes highly necessary to pinpoint a detailed relationship between different defects and coherence loss mechanisms and provide insights for ultimately optimizing the qubit's lifetime.

In this work, we introduce THz nanoimaging and nanospectroscopy as a non-destructive tool to probe materials-level sources of performance variations giving rise to decoherence in quantum circuits. By combining broadband THz pulses with an atomic force microscope (AFM), THz scattering-type scanning near-field optical microscopy (THz-sSNOM) is an ideal way to capture variations in the local conductivity spectrum at the nanoscale \cite{kim1:2023}. Recently, its capability has been demonstrated to provide reliable device defect detection in the pursuit of improved fabrication procedures for superconducting transmon qubit devices \cite{guo:2021,guo:2023,kim2:2023}. The tip-scattered THz wave detection has the advantage to observe local high-frequency conductivities without additional fabrication procedures for Ohmic contacts or destructive sample preparation techniques of the area of interest, and the individual photon energies of THz light ($hf$ = 4 meV) are several orders weaker compared with the energy range involved in high-resolution electron microscopy techniques. This makes utilizing THz-sSNOM a fast and noninvasive approach to study electrodynamics in qubit chip structures at small length scales relevant to real devices.
Note that both THz-sSNOM and AFM can achieve similar spatial resolution and serve as non-destructive, proxy techniques for examining trench depth and geometric defects. However, a key advantage of THz-sSNOM is its ability to reveal not only geometric steps but also localized "hot spots" and concentrated electric fields around sharp edges, sidewalls, and trench defects \cite{kim2:2023}—features that AFM mostly overlook. In fact, the scattering near-field signals observed via THz-sSNOM correlate well with simulated electric fields in the GHz range, because both wavelengths are significantly larger than the nanometer-scale defective features of interest.

\begin{figure*} [!ht]
\includegraphics[width=\textwidth]{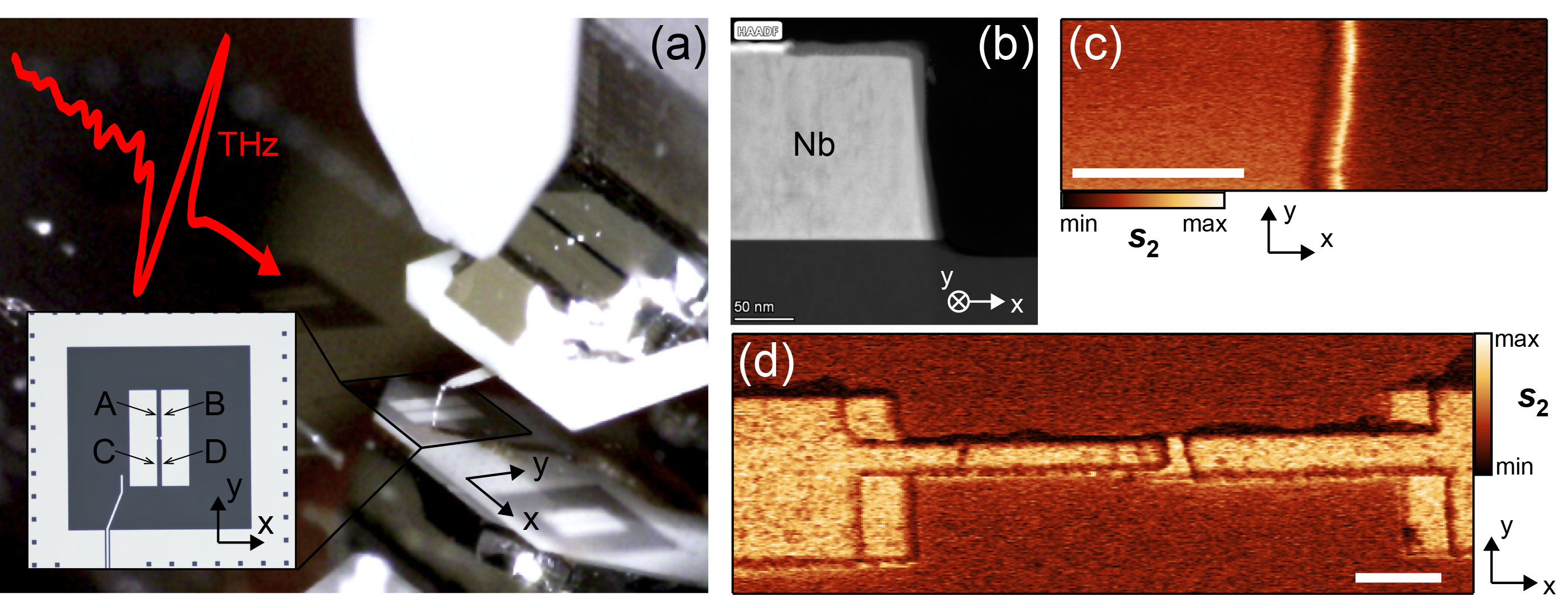}
\caption{THz nanoimaging of a qubit chip. (a) Photo of the AFM tip above the qubit sample. A THz light pulse (red) couples to the tip and results in polarizing the material underneath, producing scattered THz transients to the far field that is recorded using electro-optic sampling (not shown). Letters A, B, C, and D on the lower left inset show the locations where the Nb sidewalls were probed in a single qubit for the analysis described in Fig. \ref{fig2}. (b) Cross-sectional TEM image of a Nb sidewall. Scale bar, 50 nm. The dark area at the bottom is the sapphire substrate. The bright line on top of Nb at the left side is the AuPd capping layer. The grayish line covering the Nb at the sides and at the exposed top surface is the amorphous Nb oxide layer. THz nanoimaging of a representative (c) Nb pad edge (location C of `Qubit 3') and (d) Josephson junction area (Qubit 4). Scale bars, 1 $\mu$m.
\label{fig1}}
\end{figure*}

Our THz near-field experiment was performed on qubit chip devices prepared 
by the Superconducting Quantum Materials and Systems (SQMS) center to examine sources of loss in state-of-the-art transmon qubits \cite{center_paper}. As shown in the lower left inset of Fig. \ref{fig1}(a), the geometry of the qubit consists of a pair of rectangular shunting Nb capacitor paddles joined by a single Al/AlO$_{x}$/Al Josephson junction at the center. Qubits were prepared on substrates of double-side polished HEMEX grade HEM Sapphire. More details regarding film deposition, lithography, etching and junction fabrication parameters are provided in a previous SQMS study \cite{bal:2024}. The encapsulated qubits are fabricated by preparing 6 nm of gold palladium (AuPd) on top of a 155-nm-thick Nb. As for qubit measurements, the $T_{1}$ characterization is carried out, as it is widely believed to directly reflect the loss due to TLS. Details regarding the microwave package, microwave shielding environment and various measurement setup parameters are also provided in Ref. \citenum{bal:2024}. The qubit state was measured via dispersive readout, as detailed in Ref. \citenum{bal:2024} as well.

Broadband THz pulses are generated by using an ytterbium laser with a pulse energy of 20 $\mu$J, a repetition rate of 1 MHz, a pulse width of 91 fs, and a central laser wavelength of 1030 nm, which is incident onto a cantilever-based tapping-mode AFM system \cite{kim1:2023}. For sSNOM in general, a nominal tip radius of 20 nm determines the spatial resolution of the near-field images. Once the tip is in contact with the sample, a time-domain THz spectroscopy can be performed by moving the motorized stage that controls the time delay of the optical sampling pulse to the electro-optic crystal, tracing out the oscillating electric-field waveform of the scattered THz near-field amplitude. To obtain near-field images, the sample stage underneath the tip is raster scanned, while the THz sampling delay is fixed to a position that gives the largest near-field signal amplitude. The AFM height information is also taken simultaneously along with the near-field imaging. Near-field signals $\mathbf{s}_{n}$ are extracted from the scattered THz signal by demodulating the backscattered radiation collected from the tip–sample system at $n$th harmonics of the tip-tapping frequency from the AFM ($n$ = 1, 2, 3, and 4).
More details of our THz-sSNOM setup have been extensively discussed elsewhere \cite{n4,n5,n6}. 

Raster scans are performed with a sampling time of 90 ms per pixel and a pixel size of 10 nm and 20 nm for the Nb sidewall area (Fig. \ref{fig1}(c)) and the Al Josephson junction area (Fig. \ref{fig1}(d)), respectively. As expected, the metallic Nb and Al exhibits a higher scattered amplitude compared to the sapphire substrate. This advancing technique has recently reached a new milestone with the capability of operating at temperatures down to 1.8 K, in magnetic fields up to 5 T, and in the frequency range of 0–2 THz \cite{kim1:2023}. However, in the current THz nano-imaging measurements of the Nb resonator films, the samples act as bulk-like high reflectors for THz light. As a result, the contrast between the normal and superconducting states of Nb is relatively subtle, especially when compared to the strong signals arising from nanoscale inhomogeneities—such as sharp boundaries, edges, and defects—that concentrate the electric field \cite{kim2:2023}.

\begin{figure*} [!ht]
\includegraphics[width=\textwidth]{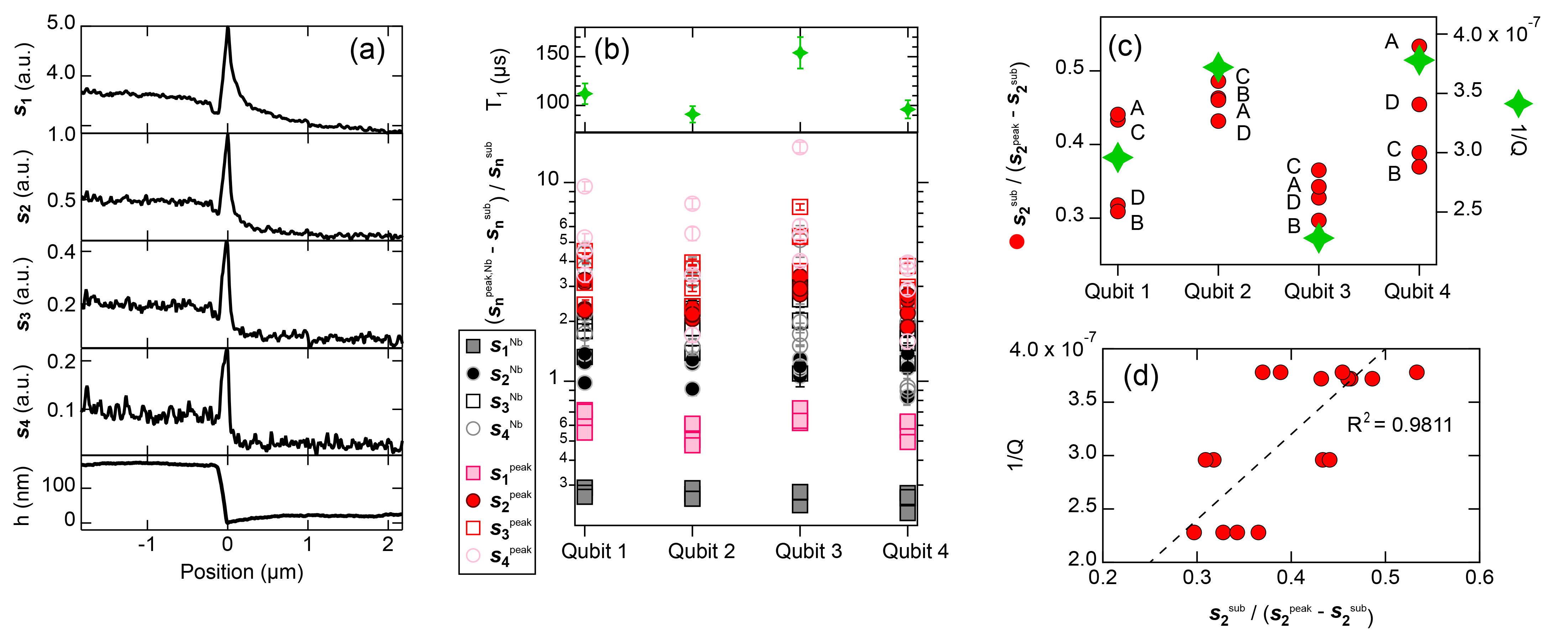}
\caption{THz near-field amplitude and qubit coherence times. (a) Line profiles across an Nb edge, such as the one presented in Fig. \ref{fig1}(c), showing the simultaneously measured near-field amplitudes $\mathbf{s}_{1}$, $\mathbf{s}_{2}$, $\mathbf{s}_{3}$, $\mathbf{s}_{4}$ and the topography taken at location B for `Qubit 3'. With respect to position at 0 $\mu$m, the left is Nb and the right is the substrate. (b) Normalized near-field signals extracted from the line profiles measured for four different qubits. In each qubit, four different locations, A, B, C, and D as indicated in the inset of Fig. \ref{fig1}(a), were probed to obtain line profiles as shown in (a), resulting in a total of 16 line cut measurements. $\mathbf{s}^{\mathrm{Nb}}_{n}$ is the value taken from the left end of the line scan and $\mathbf{s}^{\mathrm{sub}}_{n}$ is from the right end. $\mathbf{s}^{\mathrm{peak}}_{n}$ is the maximum value around 0 $\mu$m. For higher order near-field signals, the range represents the standard deviation of the noise level. The top graph shows the average $T_{1}$ values measured for each qubit. The range indicates plus and minus the standard deviation. (c) Same as graph (b) but only the second-order $\mathbf{s}^{\mathrm{peak}}_{2}$ normalized to $\mathbf{s}^{\mathrm{sub}}_{2}$ is plotted. The reciprocal values of the normalized near-field quantity and $1/Q$ instead of $T_{1}$ are plotted. Locations of where the near-field data are measured is also specified next to each symbol. (d) The near-field data in (c) re-plotted to determine the correlation with $1/Q$. 
\label{fig2}}
\end{figure*}

The encapsulation strategy currently results in a capping layer deposited only on the top surface of Nb and leaves the sidewalls exposed for surface oxides to form. The sidewall oxides can have a large participation ratio in the gap region between metal structures and at sharp corners where the electric fields tend to be concentrated. Thus, evaluating the contribution of how structural and material properties of sidewalls influence device performances will be important to eliminate losses in encapsulated qubits. The bright line along the Nb sidewall as observed in Fig. \ref{fig1}(c) is conventionally known as the edge effect in sSNOM measurements that occurs at sharp material boundaries due to a topographical crosstalk \cite{tau:2003, bab:2017}. From the fact that near-field enhancements occur around topographical boundaries, we hypothesized that the scattered amplitude at the Nb edges captures important sidewall features that can be related to qubit performances.

To this end, we conducted a set of line scans across Nb sidewalls in the gap region on four different qubits. Fig. \ref{fig2}(a) shows an example line scan of the simultaneously taken THz near-field scattered amplitude signals and the AFM height profile. Next, we compare measurements of sixteen line profiles probed at four different locations for each of the four qubits (Fig. \ref{fig2}(b)). The maximum near-field signals at the Nb edge $\mathbf{s}^{\mathrm{peak}}_{n}$, after removing the offset by the substrate signal $\mathbf{s}^{\mathrm{sub}}_{n}$ and normalizing by $\mathbf{s}^{\mathrm{sub}}_{n}$, are plotted altogether for the four samples. These near-field values interestingly follow the same trend with the average $T_{1}$ measurements of the four qubits as shown in the upper graph of Fig. \ref{fig2}(b), i.e., samples measured with a higher ($\mathbf{s}^{\mathrm{peak}}_{n}-\mathbf{s}^{\mathrm{sub}}_{n})/\mathbf{s}^{\mathrm{sub}}_{n}$ value also corresponds to having a longer $T_{1}$ time. In comparison, when we choose the scattered signal from the Nb side ($\mathbf{s}^{\mathrm{Nb}}_{n}$) and plot its normalized difference with respect to $\mathbf{s}^{\mathrm{sub}}_{n}$, we do not observe ($\mathbf{s}^{\mathrm{Nb}}_{n}-\mathbf{s}^{\mathrm{sub}}_{n})/\mathbf{s}^{\mathrm{sub}}_{n}$ having any resemblance with the $T_{1}$ values as was demonstrated with ($\mathbf{s}^{\mathrm{peak}}_{n}-\mathbf{s}^{\mathrm{sub}}_{n})/\mathbf{s}^{\mathrm{sub}}_{n}$. The relationship between the two can be examined in detail with Fig. \ref{fig2}(c) which plots just the second-order $\mathbf{s}^{\mathrm{peak}}_{2}$ data (left axis) together with $T_{1}$ (right axis). $\mathbf{s}_{2}$ is chosen since it better reflects the near-field characteristics compared to $\mathbf{s}_{1}$ and has a higher signal-to-noise ratio than $\mathbf{s}_{3}$ and $\mathbf{s}_{4}$. We use the reciprocal value $\mathbf{s}^{\mathrm{sub}}_{2}/(\mathbf{s}^{\mathrm{peak}}_{2}-\mathbf{s}^{\mathrm{sub}}_{2}$) instead of ($\mathbf{s}^{\mathrm{peak}}_{2}-\mathbf{s}^{\mathrm{sub}}_{2})/\mathbf{s}^{\mathrm{sub}}_{2}$ to match with $1/Q$, which has the useful property of being additive when comparing with other loss mechanisms contributing to the quality factor $Q$, and $Q$ is equal to the qubit frequency $f$ multiplied by the average $T_{1}$ times $2\pi$ (Qubit 1: $f$ = 4.798 GHz; Qubit 2: $f$ = 4.697 GHz; Qubit 3: $f$ = 4.530 GHz; Qubit 4: $f$ = 4.387 GHz). Also, by plotting the same results with $1/Q$ on the $y$-axis and the near-field data $\mathbf{s}^{\mathrm{sub}}_{2}/(\mathbf{s}^{\mathrm{peak}}_{2}-\mathbf{s}^{\mathrm{sub}}_{2}$) on the $x$-axis to inspect their correlation, we observe a clear linear trend between the two quantities. On the other hand, this linear relation with the near-field data is completely absent when we conducted identical measurements on four qubits prepared without encapsulation and thus exhibiting lower average $T_{1}$ times. Although experiments on more capped Nb qubits with different relaxation times would be necessary to acquire convincing statistics, we believe our results from the four samples with and without capping suggests that the exposed Nb sidewalls play an important role in influencing coherence times for transmon qubits fabricated by Nb surface encapsulation and that near-field microscopy can be used as a room-temperature measurement technique with rapid turnaround times to evaluate these qubits.

We further compared the near-field results with cross-sectional transmission electron microscopy (TEM) images of the Nb sidewall taken from the four capped qubits to assess which structural or compositional differences lead to the variation in the qubit performances. We found two structural features that appeared to provide a link between the THz near-field data and the performance variations of the transmons. As representatively shown in Fig. \ref{fig1}(b), the capping layer of AuPd stopped short before reaching the end of Nb in all the qubits which left an unprotected Nb top surface extending roughly 100 nm to oxidize, likely a result of local variations in the quantity of gold etchant across the wafer. We observe a weak trend between this exposed oxide region and decreasing quality factor. Also, we observed slight differences in the trench depth, which is the depth due to the exposed substrate area having been etched out so that it lies lower than the actual Nb and substrate interface. Interestingly, we noticed that there appears to be a relationship between the THz near-field signals from the Nb edges and trench depth for both qubits having base layers of Nb prepared with or without AuPd encapsulation \cite{center_paper}. The observed correlation between increasing trench depth and increased $Q$ for the four encapsulated qubits is in agreement with previous simulation works \cite{gam:2017,mur:2020}.  
Other features such as the oxide layer thickness or the edge curvature and slope angle of the Nb sidewall exhibited little or no noticeable difference between the qubits. Thus, although we do not observe a single defining microscopic feature that alone relates to the $T_{1}$ values, we speculate that the recorded THz near-field scattering amplitude comprehensively records the various structural factors and surface chemistry from the sidewall and can serve as a proxy to evaluate qubit coherences as elaborated below. 

\begin{figure*} [!ht]
\includegraphics[width=\textwidth]{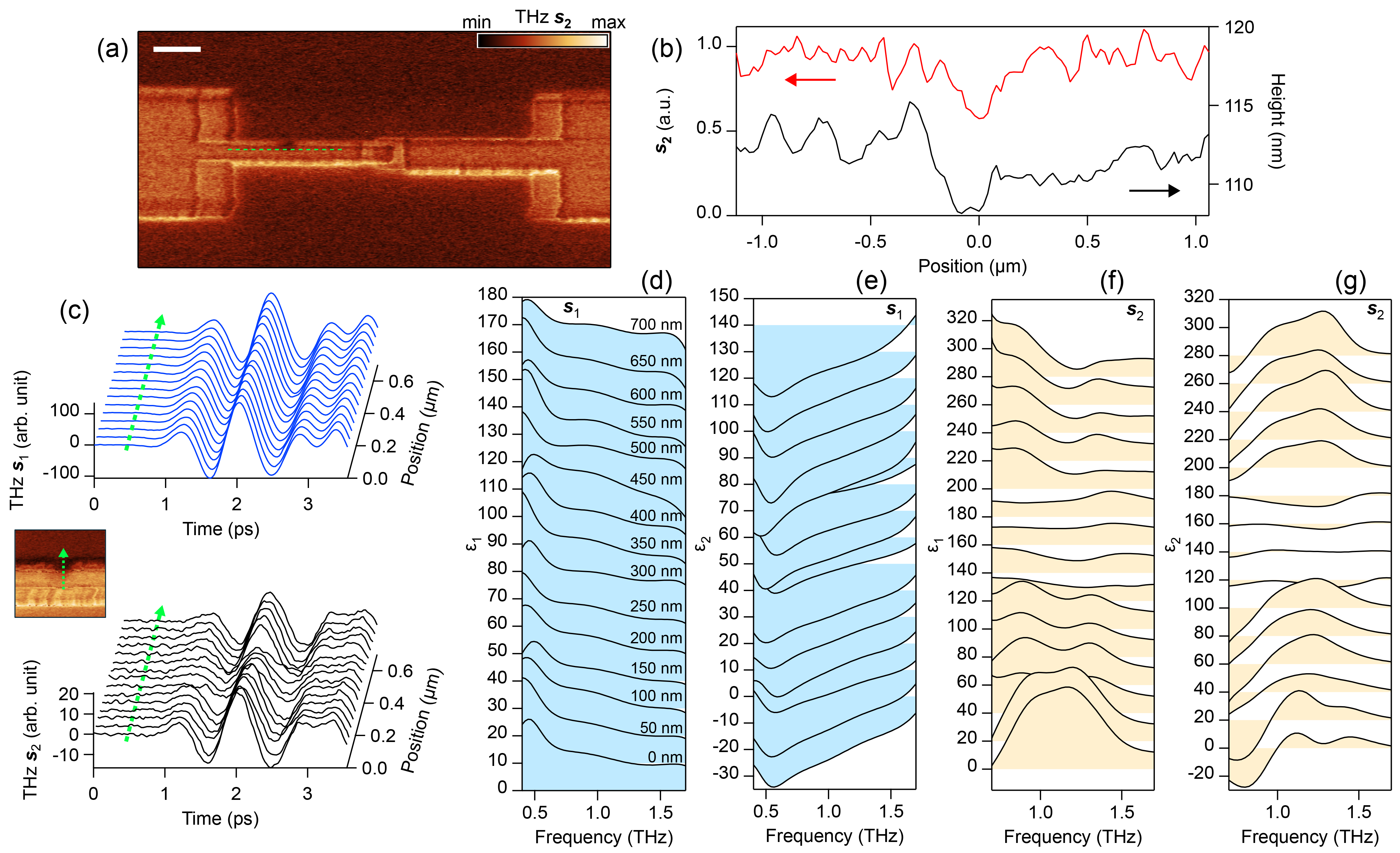}
\caption{THz nanospectroscopy across a Josephson junction structure. (a) THz $\mathbf{s}_{2}$ image of a Al Josephson junction area. Scale bar, 1 $\mu$m. (b) $\mathbf{s}_{2}$ and simultaneously measured topography line cut along the dotted green line in (a). The defect spot is centered around 0 $\mu$m. (c) $\mathbf{s}_{1}$ and $\mathbf{s}_{2}$ THz time-domain traces measured for each position separated by 50 nm steps across the dark defect at the left arm of the Al junction. (d), (f) Real and (e), (g) imaginary parts of the dielectric constants extracted from $\mathbf{s}_{1}$ and $\mathbf{s}_{2}$ THz time traces, respectively. Spectra are offset for clarity.
\label{fig4}}
\end{figure*}

Near-field scattering signals are exceptionally sensitive to abrupt changes in geometry or material properties. In particular, the discontinuities and imperfections along the trench and sidewalls at the sapphire/Nb and Nb/air interfaces act as strong scattering centers. Under the near-field tip, these geometric features give rise to local electric field enhancements, owing to their high polarizability. The intensified local field then polarizes the tip antenna itself, which in turn radiates a scattered signal detectable by the electro-optical detector.
Electromagnetic analysis of a transmon qubit operating at 5 GHz shows that changes in the sidewall angle and trench depth measured between multiple qubits can affect the energy participation ratio (EPR) of these regions by a factor of 1.3 and 2, respectively \cite{center_paper}. Although the electromagnetic modes at THz and GHz frequencies differ, both fall into the electrostatic regime in this context because their wavelengths are much larger than the nanometer-scale trench and sidewall features.

In addition to the geometric imperfections, near-field scattering signals can also originate from inhomogeneities in the Nb film and its oxide layers. However, such materials-based variations typically induce much smaller field enhancements compared to the dominant contributions from the trench and sidewall geometries.
Both the device sidewall of the antenna pads and the Josephson junction areas significantly contribute to the overall electrical participation ratio. However, our electrostatic simulation indicates that the front portion of the antenna pads contributes more than twice as much as the JJ area alone \cite{center_paper}. This conclusion aligns with the observed stronger correlation between the Q-factor and sidewall imperfections.
In addition, we anticipate that extending the current GHz electrostatic simulations into the THz regime and linking THz electrodynamics with sSNOM measurements of the qubit circuit will be necessary to quantitatively substantiate this correlation. Nevertheless, we expect qualitatively similar electrostatic results, which lie beyond the scope of our present study.

Lastly, we demonstrate the capability of THz nanoscopy to find a structural defect in one of the Al Josephson junctions and to extract dielectric functions in the GHz/THz frequency range with nanometer spatial precision. The Dolan bridge style Al/AlO$_{x}$/Al Josephson junctions are fabricated using the well-known double angle shadow evaporation technique and solvent lift off \cite{bal:2024}. The Josephson junction is defined by the region where an upper Al lead overlaps a lower lead separated by an insulating oxide layer shown at the center of the THz near-field image in Fig. \ref{fig4}(a). A dark spot is clearly observed in the left arm of the junction in this image. A close inspection using a line cut across this spot (Fig. \ref{fig4}(b)) shows a slight indentation in height of less than 5 nm where the THz $\mathbf{s}_{2}$ drops about $30\sim40\%$. This behavior is in contrast to what is normally observed in a topographical depression where the SNOM tip can have an enhanced interaction with the material surroundings at this recess and translates to an increased near-field signal. To detect the cause of the sudden drop in near-field scattered amplitude, we conducted near-field time-domain THz spectroscopy at every steps of 50 nm acoss the dark spot starting from the Al lead wire and ending in the substrate (Fig. \ref{fig4}(c)). The complex dielectric constants for each position is then retrieved from both $\mathbf{s}_{1}$ (Fig. \ref{fig4}(d) and (e)) and $\mathbf{s}_{2}$ (Fig. \ref{fig4}(f) and (g)) time trace measurements \cite{guo2:2021}. We note that these are effective dielectric constants reflecting how the THz electromagnetic fields are responding to the nanostructures of the junction. The subwavelength structural feature renders a flow of free carriers unlike what is described for light-matter interaction in a bulk material. The spectra not showing a good resemblance with electrodynamics expected from conventional materials can be explained by this metamaterial-like behavior from the junction structure. While the dielectric response measured from $\mathbf{s}_{1}$ does not exhibit a noticeable difference within the sub-micrometer scan, $\epsilon_{1}$ and $\epsilon_{2}$ from $\mathbf{s}_{2}$ clearly resolve the nanoscale near-field response and display distinctive contrast between the Al part (bottom six traces) and substrate area (top five traces) and the dark spot in the middle. 

At the current stage, a fully quantitative simulation of the observed local THz electrodynamics remains challenging. Nevertheless, we note that many of the measured $\mathbf{s}_{2}$ dielectric spectra exhibit suppressed conductivity at lower frequencies, indicative of a Drude-Smith model capturing carrier confinement effects in the junction from GHz to THz frequencies. Moreover, the observed correlation between local electrodynamics and device performance is weaker in nano-junction areas compared to the sidewall scattering illustrated in Fig.\ref{fig2}(c). This observation motivates further advancements in ultra-low cryogenic sSNOM beyond the state of art \cite{kim1:2023, sam} to achieve milliKelvin temperatures below the aluminum junction critical temperature. Additionally, while the presence of a local defect in the junction (Fig.\ref{fig4}(a)) did not directly correlate with decreased qubit $T_{1}$ times, the demonstrated capability to probe local THz electrodynamics at spatial scales of approximately 20 nm (Figs.\ref{fig4}(c)-(g)) is expected to significantly impact future experiments exploring collective modes \cite{n1, n3} and topological phases \cite{n2} in quantum materials and devices.


In conclusion, we employed THz near-field microscopy to probe materials defects in quantum circuits and discovered a potential figure of merit that can assist in evaluating the coherence times of individual superconducting qubits. This method provides a noninvasive room-temperature screening to easily examine the sidewall characteristics in a comprehensive fashion for encapsulated transmon devices. With THz nanoscopy extended to cryogenic temperatures \cite{kim1:2023}, we anticipate nanoimaging and nanospectroscopy in the GHz to THz frequency range to reveal electromagnetic field distributions relating to local charge and current distributions from Cooper pairs and quasiparticles which will offer new insights in assessing the designs and noise properties of superconducting quantum systems at the nanoscale.  

\begin{acknowledgments}
The THz-sSNOM sharacterization and Fabrication of transmon circuits were supported
by the U.S. Department of Energy, Office of Science, National Quantum Information Science
Research Centers, Superconducting Quantum Materials and Systems Center (SQMS) under
the contract No. DEAC02-07CH11359. 
Modeling were supported by the U.S. Department of Energy, Office of Basic Energy Science, Division of Materials
Sciences and Engineering (Ames National Laboratory is operated for the U.S. Department of
Energy by Iowa State University under Contract No. DE-AC02-07CH11358).

\end{acknowledgments}

\section*{Data Availability Statement}

All data supporting the conclusions of this paper are included within the paper and/or available upon request from J.W.

\nocite{*}

\begin{thebibliography}{1}

\bibitem{ab:2024}
B. Abdisatarov, D. Bafia, A. Murthy, G. Eremeev, H. E. Elsayed-Ali, J. Lee, A. Netepenko, C. P. A. Carlos, S. Leith, G. J. Rosaz, A. Romanenko, and A. Grassellino, “Direct measurement of microwave loss in Nb films for superconducting qubits,” \href{https://doi.org/10.1063/5.0226611}{Appl. Phys. Lett.} \textbf{125}, 124002 (2024).

\bibitem{baf:2024}
D. Bafia, A. Murthy, A. Grassellino, and A. Romanenko, “Oxygen vacancies in niobium pentoxide as a source of two-level system losses in superconducting niobium,” \href{https://doi.org/10.1103/PhysRevApplied.22.024035}{Phys. Rev. Appl.} \textbf{22}, 024035 (2024).

\bibitem{oh1:2024}
J.-S. Oh, C. J. Kopas, J. Marshall, X. Fang, K. R. Joshi, A. Datta, S. Ghimire, J.-M. Park, R. Kim, D. Setiawan, E. Lachman, J. Y. Mutus, A. A. Murthy, A. Grassellino, A. Romanenko, J. Zasadzinski, J. Wang, R. Prozorov, K. Yadavalli, M. Kramer, and L. Zhou, “Exploring the relationship between deposition method, microstructure, and performance of Nb/Si-based superconducting coplanar waveguide resonators,” \href{https://doi.org/10.1016/j.actamat.2024.120153}{Acta Mater.} \textbf{276}, 120153 (2024).

\bibitem{oh2:2024}
J.-S. Oh, R. Zaman, A. A. Murthy, M. Bal, F. Crisa, S. Zhu, C. G. Torres-Castanedo, C. J. Kopas, J. Y. Mutus, D. Jing, J. Zasadzinski, A. Grassellino, A. Romanenko, M. C. Hersam, M. J. Bedzyk, M. Kramer, B.-C. Zhou, and L. Zhou, “Structure and Formation Mechanisms in Tantalum and Niobium Oxides in Superconducting Quantum Circuits,” \href{https://doi.org/10.1021/acsnano.4c05251}{ACS Nano} \textbf{18}, 19732-19741 (2024).

\bibitem{n7}
J.-M. Park, Z. X. Chong, R. H. J. Kim, S. Haeuser, R. Chan, A. A. Murthy, C. J. Kopas, J. Marshall, D. Setiawan, E. Lachman, J. Y. Mutus, K. Yadavalli, A. Grassellino, A. Romanenko, and J. Wang, “Probing Non-Equilibrium Pair-Breaking and Quasiparticle Dynamics in Nb Superconducting Resonators Under Magnetic Fields,” \href{https://doi.org/10.3390/ma18030569}{Materials} \textbf{18}, 569 (2025).

\bibitem{bal:2024}
M. Bal, A. A. Murthy, S. Zhu, F. Crisa, X. You, Z. Huang, T. Roy, J. Lee, D. van Zanten, R. Pilipenko, I. Nekrashevich, A. Lunin, D. Bafia, Y. Krasnikova, C. J. Kopas, E. O. Lachman, D. Miller, J. Y. Mutus, M. J. Reagor, H. Cansizoglu, J. Marshall, D. P. Pappas, K. Vu, K. Yadavalli, J.-S Oh, L. Zhou, M. J. Kramer, F. Lecocq, D. P. Goronzy, C. G. Torres-Castanedo, P. G. Pritchard, V. P. Dravid, J. M. Rondinelli, M. J. Bedzyk, M. C. Hersam, J. Zasadzinski, J. Koch, J. A. Sauls, A. Romanenko, and A. Grassellino, “Systematic improvements in transmon qubit coherence enabled by niobium surface encapsulation,” \href{https://doi.org/10.1038/s41534-024-00840-x}{npj Quantum Inf.} \textbf{10}, 43 (2024).

\bibitem{kim1:2023}
R. H. J. Kim, J.-M. Park, S. J. Haeuser, L. Luo, and J. Wang, “A sub-2 Kelvin cryogenic magneto-terahertz scattering-type scanning near-field optical microscope (cm-THz-sSNOM),” \href{https://doi.org/10.1063/5.0130680}{Rev. Sci. Instrum.} \textbf{94}, 043702 (2023).

\bibitem{guo:2021}
X. Guo, X. He, Z. Degnan, B. C. Donose, K. Bertling, A. Fedorov, A. D. Raki\'{c}, and P. Jacobson, “Near-field terahertz nanoscopy of coplanar microwave resonators,” \href{https://doi.org/10.1063/5.0061078}{Appl. Phys. Lett.} \textbf{119}, 091101 (2021).

\bibitem{guo:2023}
X. Guo, Z. Degnan, J. A. Steele, E. Solano, B. C. Donose, K. Bertling, A. Fedorov, A. D. Raki\'{c}, and P. Jacobson, “Near-Field Localization of the Boson Peak on Tantalum Films for Superconducting Quantum Devices,” \href{https://doi.org/10.1021/acs.jpclett.3c00850}{J. Phys. Chem. Lett.} \textbf{14}, 4892-4900 (2023).

\bibitem{kim2:2023}
R. H. J. Kim, J. M. Park, S. Haeuser, C. Huang, D. Cheng, T. Koschny, J. Oh, C. Kopas, H. Cansizoglu, K. Yadavalli, J. Mutus, L. Zhou, L. Luo, M. J. Kramer, and J. Wang, “Visualizing heterogeneous dipole fields by terahertz light coupling in individual nano-junctions,” \href{https://doi.org/10.1038/s42005-023-01259-0}{Commun. Phys.} \textbf{6}, 147 (2023).

\bibitem{center_paper}
A. A. Murthy, M. Bal, M. J. Bedzyk, H. Cansizoglu, R. K. Chan, V. Chandrasekhar, F. Crisa, A. Datta, Y. Deng, C. D. M. Diaz, V. P. Dravid, D. A. Garcia-Wetten, S. Ghimire, D. P. Goronzy, S. de Graaf, S. Haeuser, M. C. Hersam, P. Hopkins, D. Isheim, K. Joshi, R. Kim, C. J. Kopas, M. J. Kramer, E. O. Lachman, J. Lee, P. G. Lim, A. Lunin, W. Mah, J. Marshall, J. Y. Mutus, J.-S. Oh, D. Olaya, D. P. Pappas, J.-M. Park, R. Prozorov, R. do Reis, D. Seidman, Z. Sung, M. Tanatar, M. J. Walker, J. Wang, L. Zhou, S. Zhu, A. Grassellino, and A. Romanenko, “Identifying Materials-Level Sources of Performance Variation in Superconducting Transmon Qubits,” arXiv:2503.14424 (2025). 

\bibitem{n4} R. H. J. Kim, C. Huang, Y. Luan, L.-L. Wang, Z. Liu, J.-M. Park, L. Luo, P. M. Lozano, G. Gu, D. Turan, N. T. Yardimci, M. Jarrahi, I. E. Perakis, Z. Fei, Q. Li, and J. Wang. “Terahertz Nano-Imaging of Electronic Strip Heterogeneity in a Dirac Semimetal,” \href{https://doi.org/10.1021/acsphotonics.1c00216}{ACS Photonics} \textbf{8}, 1873-1880 (2021).

\bibitem{n5} R. H. J. Kim, Z. Liu, C. Huang, J.-M. Park, S. J. Haeuser, Z. Song, Y. Yan, Y. Yao, L. Luo, and J. Wang, “Terahertz Nanoimaging of Perovskite Solar Cell Materials,” \href{https://10.1021/acsphotonics.2c00861}{ACS Photonics} \textbf{9}, 3550-3556 (2022).

\bibitem{n6} R. H. J. Kim, A. K. Pathak, J.-M. Park, M. Imran, S. J. Haeuser, Z. Fei, Y. Mudryk, T. Koschny, and J. Wang, “Nano-compositional imaging of the lanthanum silicide system at THz wavelengths,” \href{https://10.1021/10.1364/OE.507414}{Opt. Express} \textbf{32}, 2356-2363 (2024).

\bibitem{tau:2003}
T. Taubner, R. Hillenbrand, and F. Keilmann, “Performance of visible and mid-infrared scattering-type near-field optical microscopes,” \href{https://doi.org/10.1046/j.1365-2818.2003.01164.x}{J. Microsc.} \textbf{210}, 311-314 (2003).

\bibitem{bab:2017}
V. E. Babicheva, S. Gamage, M. I. Stockman, and Y. Abate, “Near-field edge fringes at sharp material boundaries,” \href{https://doi.org/10.1364/OE.25.023935}{Opt. Express} \textbf{25}, 23935-23944 (2017).

\bibitem{gam:2017}
J. M. Gambetta, C. E. Murray, Y.-K.-K. Fung, D. T. McClure, O. Dial, and W. Shanks, “Investigating Surface Loss Effects in Superconducting Transmon Qubits,” \href{https://doi.org/10.1109/TASC.2016.2629670}{IEEE Trans. Appl. Supercond.} \textbf{27}, 1700205 (2017).

\bibitem{mur:2020}
C. E. Murray, “Analytical Modeling of Participation Reduction in Superconducting Coplanar Resonator and Qubit Designs Through Substrate Trenching,” \href{https://doi.org/10.1109/TMTT.2020.2995894}{IEEE Trans. Microw. Theory Techn.} \textbf{68}, 3263-3270 (2020).

\bibitem{guo2:2021}
X. Guo, K. Bertling, and A. D. Raki\'{c}, “Optical constants from scattering-type scanning near-field optical microscope,” \href{https://doi.org/10.1063/5.0036872}{Appl. Phys. Lett.} \textbf{118}, 041103 (2021).

\bibitem{sam}
S. Haeuser, R. H. J. Kim, J.-M. Park, R. K. Chan, M. Imran, T. Koschny, and J. Wang  “Analysis of Near-Field Magnetic Responses on ZrTe$_{5}$ through Cryogenic Magneto-THz Nano-Imaging,” \href{https://doi.org/10.3390/instruments8010021}{Instruments} \textbf{8}, 21 (2024).

\bibitem{n1} C. Huang, M. Mootz, L. Luo, D. Cheng, A. Khatri, J.-M. Park, R. H. J. Kim, Y. Qiang, V. L. Quito, Y. Yao, P. P. Orth, I. E. Perakis, and J. Wang, “Discovery of an Unconventional Quantum Echo by Interference of Higgs Coherence,” \href{
https://doi.org/10.48550/arXiv.2312.10912}{Sci. Adv.} \textbf{in press}, arXiv:2312.10912 (2025).

\bibitem{n3} L. Luo, M. Mootz, J. H. Kang, C. Huang, K. Eom, J. W. Lee, C. Vaswani, Y. G. Collantes, E. E. Hellstrom, I. E. Perakis, C. B. Eom, and J. Wang, “Quantum coherence tomography of light-controlled superconductivity,” \href{https://doi.org/10.1038/s41567-022-01827-1}{Nat. Phys.} \textbf{19}, 201-209 (2023).

\bibitem{n2} X. Yang, C. Vaswani, C. Sundahl, M. Mootz, P. Gagel, L. Luo, J. H. Kang, P. P. Orth, I. E. Perakis, C. B. Eom, and J. Wang. “Terahertz-light quantum tuning of a metastable emergent phase hidden by superconductivity,” \href{https://doi.org/10.1038/s41563-018-0096-3}{Nat. Mater.} \textbf{17}, 586-591 (2018).

\end{thebibliography}

\end{document}